\documentclass[letter,scriptaddress,twocolumn, prl,showpacs,superscriptaddress] {revtex4-1}

	\usepackage{amsmath}%,amssymb} 
	\usepackage{makeidx}
	\usepackage{amsfonts}
	\usepackage[ansinew]{inputenc}
	\usepackage[usenames,dvipsnames]{pstricks}
	\usepackage{subfigure}
	\usepackage{epsfig}
	\usepackage{pst-grad} % For gradients
	\usepackage{pst-plot} % For axes
	\usepackage[hyperindex]{hyperref}
	\usepackage{color}

\makeindex

\begin{document}

\title{Large magnetoresistance by Pauli blockade in hydrogenated graphene}

\author{J. Guillemette}
\affiliation{Department of Physics, McGill University, Montr\'eal, Qu\'ebec, H3A 2A7, Canada}
\affiliation{Department of Physics, John Abbott College, Montr\'eal, Qu\'ebec, Canada, H9X 3L9}

\author{N. Hemsworth}
\affiliation{Department of Electrical and Computer Engineering, McGill University, Montr\'eal, Qu\'ebec, H3A 2A7, Canada}

\author{A. Vlasov}
\affiliation{Department of Electrical and Computer Engineering, McGill University, Montr\'eal, Qu\'ebec, H3A 2A7, Canada}

\author{J. Kirman}
\affiliation{Department of Electrical and Computer Engineering, McGill University, Montr\'eal, Qu\'ebec, H3A 2A7, Canada}

\author{F. Mahvash}
\affiliation{Department of Electrical and Computer Engineering, McGill University, Montr\'eal, Qu\'ebec, H3A 2A7, Canada}
\affiliation{Department of Chemistry, Universit\'e du Qu\'ebec \`a Montr\'eal, Montr\'eal, Qu\'ebec, H3C 3P8, Canada}

\author{P. L. L\'evesque}
\affiliation{Department of Chemistry, Universit\'e de Montr\'eal, Montr\'eal, Qu\'ebec, H3C 3J7, Canada}

\author{M. Siaj }
\affiliation{Department of Chemistry, Universit\'e du Qu\'ebec \`a Montr\'eal, Montr\'eal, Qu\'ebec, H3C 3P8, Canada}

\author{R. Martel }
\affiliation{Department of Chemistry, Universit\'e de Montr\'eal, Montr\'eal, Qu\'ebec, H3C 3J7, Canada}

\author{G. Gervais }
\affiliation{Department of Physics, McGill University, Montr\'eal, Qu\'ebec, H3A 2A7, Canada}

\author{S. Studenikin }
\affiliation{National Research Council Canada, 1200 Montreal Road, Ottawa, Ontario, K1A 0R6}

\author{A. Sachrajda }
\affiliation{National Research Council Canada, 1200 Montreal Road, Ottawa, Ontario, K1A 0R6}

\author{T. Szkopek}
\email{thomas.szkopek@mcgill.ca}
\affiliation{Department of Electrical and Computer Engineering, McGill University, Montr\'eal, Qu\'ebec, H3A 2A7, Canada}

\begin{abstract}
We report the observation of a giant positive magnetoresistance in millimetre scale hydrogenated graphene with magnetic field oriented in the plane of the graphene sheet. A positive magnetoresistance in excess of 200\% at a temperature of 300 mK was observed in this configuration, reverting to negative magnetoresistance with the magnetic field oriented normal to the graphene plane. We attribute the observed positive, in-plane, magnetoresistance to Pauli-blockade of hopping conduction induced by spin polarization. Our work shows that spin polarization in concert with electron-electron interaction can play a dominant role in magnetotransport within an atomic monolayer.
\end{abstract}

%\pacs{65.80.Ck}

\maketitle

Giant magnetoresistance (GMR) is a manifestation of spin dependent charge transport that encompasses a wide range of phenomena \cite{fert, grunberg}. The strength of the GMR effect has led to its application in the sensing of magnetic fields, most importantly in high density magnetic information storage. Here, we report the experimental discovery of a positive, in-plane magnetoresistance (MR) reaching $235\%$ in hydrogenated graphene, which we term large magnetoresistance (LMR). Similar hydrogenated graphene samples have shown a strong \textit{negative} MR with magnetic field applied normal to the sample surface, corresponding to a transition from an insulating state to a quantum Hall state \cite{guillemette13,bennaceur}. The MR of a variety of functionalized graphene systems with magnetic field applied normal to the graphene surface has been explored \cite{hong,matis,datta}. In contrast, we have observed a strong \textit{positive} magnetoresistance with the magnetic field applied in-plane. Considering the hopping conduction mechanism of hydrogenated graphene, it is the combination of spin polarization and electron-electron interaction that leads to positive LMR by Pauli blockade of electron hopping, as first described theoretically in the seminal work of Kamimura \textit{et al.}\cite{kamimura}.

Notably, the \textit{quasi}-two-dimensional nature of a 2D electron system (2DES) can itself lead to a strong positive MR with the magnetic field applied in-plane due to the finite layer thickness of the electron system. Non-perturbative magneto-orbital coupling arises when the magnetic length $\ell_B = \sqrt{h/eB_{\|}}$ approaches, and becomes smaller than, the rms thickness $\sqrt{< z^2 >}$ of the 2DES \cite{SDS00}. Experimentally, positive in-plane MR has been observed in high-mobility Si field effect transistors \cite{pudalov, simonian97, mertes99}, high mobility GaAs/AlGaAs heterostructures and quantum wells \cite{simmons98, yoon00, zhou10}, and Mn$^{+2}$ doped II-VI quantum wells \cite{smorchkova98}. Singlet and triplet correlations in a high-mobility, low-density 2DES have also been predicted to contribute to in-plane MR \cite{zala, spivak}. In-plane MR has been used as an experimental probe to gain insight into electron-electron interactions in the vicinity of the metal-insulator transition \cite{anissimova}. 

In the case of a graphene monolayer, the 2DES is confined to an atomic length scale of $\sim0.3~\mathrm{nm}$, which is orders of magnitude thinner than a typical 2DES hosted in a semiconductor heterostructure, leading to suppression of magneto-orbital effects. The resistivity of a pristine graphene monolayer encapsulated in boron nitride was measured with an in-plane magnetic field of up to $B_{\|}=30~\mathrm{T}$, corresponding to a magnetic length as short as $\ell_B \sim 4.7~\mathrm{nm}$; in-plane MR was observed to be absent to within experimental error \cite{chiappini16}. A weak in-plane MR of $\sim 3\%$ was observed in a graphene monolayer on a SiO$_2$ substrate as a result of magneto-orbital coupling through graphene ripples \cite{lundeberg}. More recently, a positive in-plane MR of $\sim5\%$ was observed in monolayer graphene on SiC and attributed to spin dependent scattering at grain boundaries \cite{wu17}. In-plane MR effects in graphene have thus far been observed to be comparatively small. We report here the observation of in-plane LMR in hydrogenated graphene, where atomic thickness suppresses magneto-orbital MR, and the primary effect of in-plane magnetic field is to suppress hopping conduction via Pauli blockade as first theoretically described by Kamimura \textit{et al.}\cite{kamimura}.

We conducted our experiments on millimetre scale samples of monolayer graphene grown by chemical vapour deposition (CVD) \cite{guermoune11} and transferred to degenerately doped silicon substrates with a 300~nm layer of oxide. Electrical contacts were fabricated either by direct mechanical application of In (sample HGT2) or by vacuum deposition of Ti/Au (3nm/50nm) through a shadow mask (samples HG18, HG55). Hydrogenation was performed in an ultra-high vacuum chamber with an atomic hydrogen beam produced by thermal cracking of molecular hydrogen in a tungsten capillary heated by electron beam bombardment \cite{guillemette13}. Atomic hydrogen adsorbates create C-H bonds that disrupt the $sp^2$ lattice of graphene to create localized $sp^3$ distortions, with a profound effect on graphene's electronic properties \cite{elias, son16}. The neutral point defect density per carbon atom induced by hydrogenation in our samples is on the order of parts per thousand, as inferred from Raman spectroscopy \cite{guillemette13, bennaceur, hemsworth}. Direct experimental evidence for band gap opening and the appearance of localized states in hydrogenated graphene has reported using angle-resolved photoemission spectroscopy (ARPES) and scanning tunnelling spectroscopy \cite{haberer1, haberer2, scheffler, goler}. 

\begin{figure}%[htbp]
%	\begin{center}
		\includegraphics[width=0.45\textwidth]{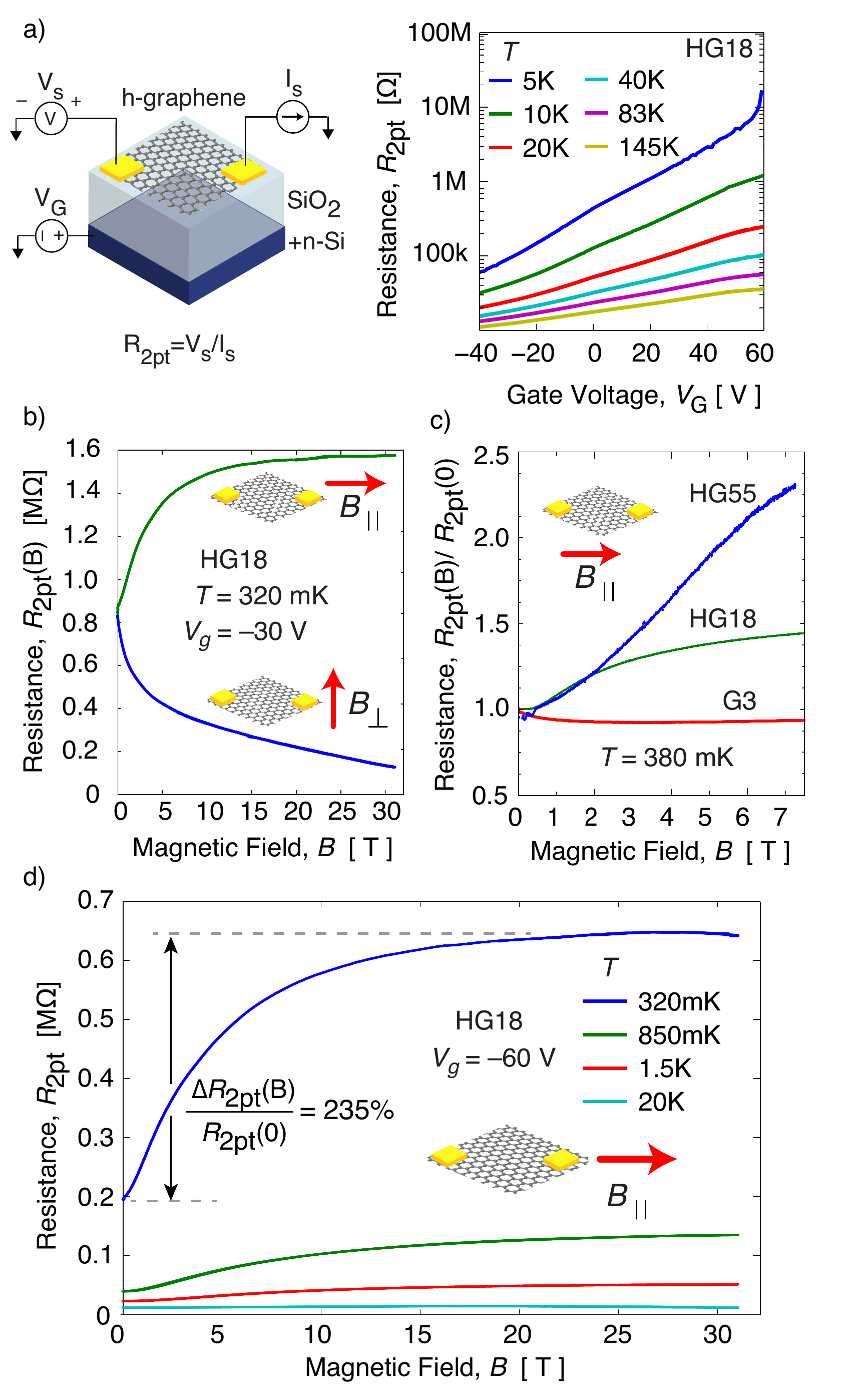} 
		\caption{ The measured 2-point resistance $R_{\mathrm{2pt}}$ of hydrogenated graphene sample HG18 at zero magnetic field versus back-gate voltage $V_G$ is plotted in a) at different temperatures $T$, showing strong insulating behaviour and hole conduction. b) The resistance $R_{\mathrm{2pt}}(B)$ of sample HG18 is plotted versus magnetic field $B$ oriented in-plane, $B_\|$, and out-of-plane, $B_\bot$, to the graphene at $T=320~\mathrm{mK}$ and $V_G = -30~\mathrm{V}$.  c) Comparison of the normalized resistance $R_{\mathrm{2pt}}(B)/R_{\mathrm{2pt}}(0)$ versus in-plane magnetic field $B$ of hydrogenated samples HG18 and HG55, with the pristine graphene sample G3. d) The temperature dependence of the resistance $R_{\mathrm{2pt}}(B)$ of sample HG18 is plotted versus in-plane magnetic field $B$ at $V_G = -60~\mathrm{V}$. A maximum magnetoresistance of $\Delta R_{\mathrm{2pt}}(B)/R_{\mathrm{2pt}}(0) = [ R_{\mathrm{2pt}}(B) -  R_{\mathrm{2pt}}(0) ] /R_{\mathrm{2pt}}(0) = 235\%$ is observed. }
		\label{fig:Thermo}
%	\end{center}
\end{figure}

We measured the electronic transport properties of hydrogenated graphene in the absence of a magnetic field, using standard lock-in detection techniques at a frequency $f\sim10~\mathrm{Hz}$. The 2-point resistance $R_{\mathrm{2pt}}$ versus gate voltage $V_\mathrm{G}$ at temperatures $T = 5-145~\mathrm{K}$ are shown in Fig. 1a) for the representative sample HG18. Strong insulating behaviour, $\partial R_{\mathrm{2pt}} / \partial T < 0 $, is observed, indicative of the onset of electron localization by the introduction of neutral point defects into the graphene lattice via hydrogenation, as previously reported \cite{elias, guillemette13, bennaceur, hemsworth}. The field effect corresponds to hole conduction, $\partial R_{\mathrm{2pt}} / \partial V_{\mathrm{G}} > 0 $, with a field effect mobility $\mu \propto \partial (1/R) / \partial V_\mathrm{G} \rightarrow 0$ as $T \rightarrow 0$. The back-gate capacitance of $C_{ox} = 11.5~\mathrm{nF/cm}^2$, leading to a modulation in hole density of $\Delta p = C_{ox} V_{\mathrm{G}} / e = 7.18\times10^{12}/\mathrm{cm}^2$ over the swept gate voltage range $V_\mathrm{G} = 100~\mathrm{V}$.

The MR of hydrogenated graphene sample HG18 was measured in a 32~T resistive magnet. The sample was mounted on a rotating sample mount to enable measurements with magnetic field $B$ applied in-plane and out-of-plane relative to the graphene. The 2-point resistance $R_{\mathrm{2pt}}$ versus in-plane and out-of-plane magnetic field $B$ is shown in Fig. 1b) at a constant gate voltage $V_\mathrm{G}=-30~\mathrm{V}$ and temperature $T = 320~\mathrm{mK}$. Strong negative MR is observed with magnetic field applied out-of-plane, as previously reported \cite{guillemette13}. In contrast, strong \textit{positive} MR is observed with magnetic field oriented in-plane.  The normalized MR is defined as $\Delta R_{\mathrm{2pt}}(B)/R_{\mathrm{2pt}}(0) = [ R_{\mathrm{2pt}}(B) -  R_{\mathrm{2pt}}(0) ] /R_{\mathrm{2pt}}(0)$, and reaches a value of $84\%$ at $B = 32~\mathrm{T}$. We ascribe the observed magnetoresistance to hydrogenation, as opposed to the native disorder of CVD grown graphene. A direct comparison of the normalized resistance $R_{\mathrm{2pt}}(B)/R_{\mathrm{2pt}}(0)$ versus in-plane magnetic field $B$ for two hydrogenated samples HG18 and HG55, and a pristine graphene sample G3, is shown in Fig. 1(c). There is no positive MR observed in pristine graphene, in agreement with Chiappini \textit{et al.} \cite{chiappini16}.  We attribute the small negative MR observed with G3 to a residual out-of-plane magnetic field component. For example, a misalignment between in-plane field and the graphene plane of $0.6^\circ$ at $B = 5~\mathrm{T}$ results in a 50~mT out-of-plane field that is sufficient to induce a negative MR of several percent due to weak localization \cite{baker}. 

The temperature dependence of the measured 2-point resistance $R_{\mathrm{2pt}}$ versus in-plane magnetic field $B_{\|}$ at a gate voltage $V_\mathrm{G}=-60~\mathrm{V}$ is shown in Fig. 1d). A positive MR reaching $\Delta R_{\mathrm{2pt}}(B)/R_{\mathrm{2pt}}(0) =235\%$ was observed at $T = 320~\mathrm{mK}$. The positive LMR is in stark contrast with the comparatively weak positive MR of up to $5\%$ reported in graphene samples under similar experimental conditions \cite{lundeberg, chiappini16, wu17}. Both the resistance and the MR are strongly suppressed as temperature increases, with the MR dropping to $\Delta R_{\mathrm{2pt}}(B)/R_{\mathrm{2pt}}(0) =125\%$ at $T=1.5~\mathrm{K}$, and $\Delta R_{\mathrm{2pt}}(B)/R_{\mathrm{2pt}}(0) = 24\%$ at $T=20~\mathrm{K}$. %The MR at more positive gate voltages ( $V_{\mathrm{G}} > -30~\mathrm{V}$ ) could not be investigated due to a combination of excessive sample resistance ($R_{\mathrm{2pt}}>2~\mathrm{M}\Omega$) and parasitic, shunt capacitance.

\begin{figure}%[htbp]
	\begin{center}
		\includegraphics[width=0.45\textwidth]{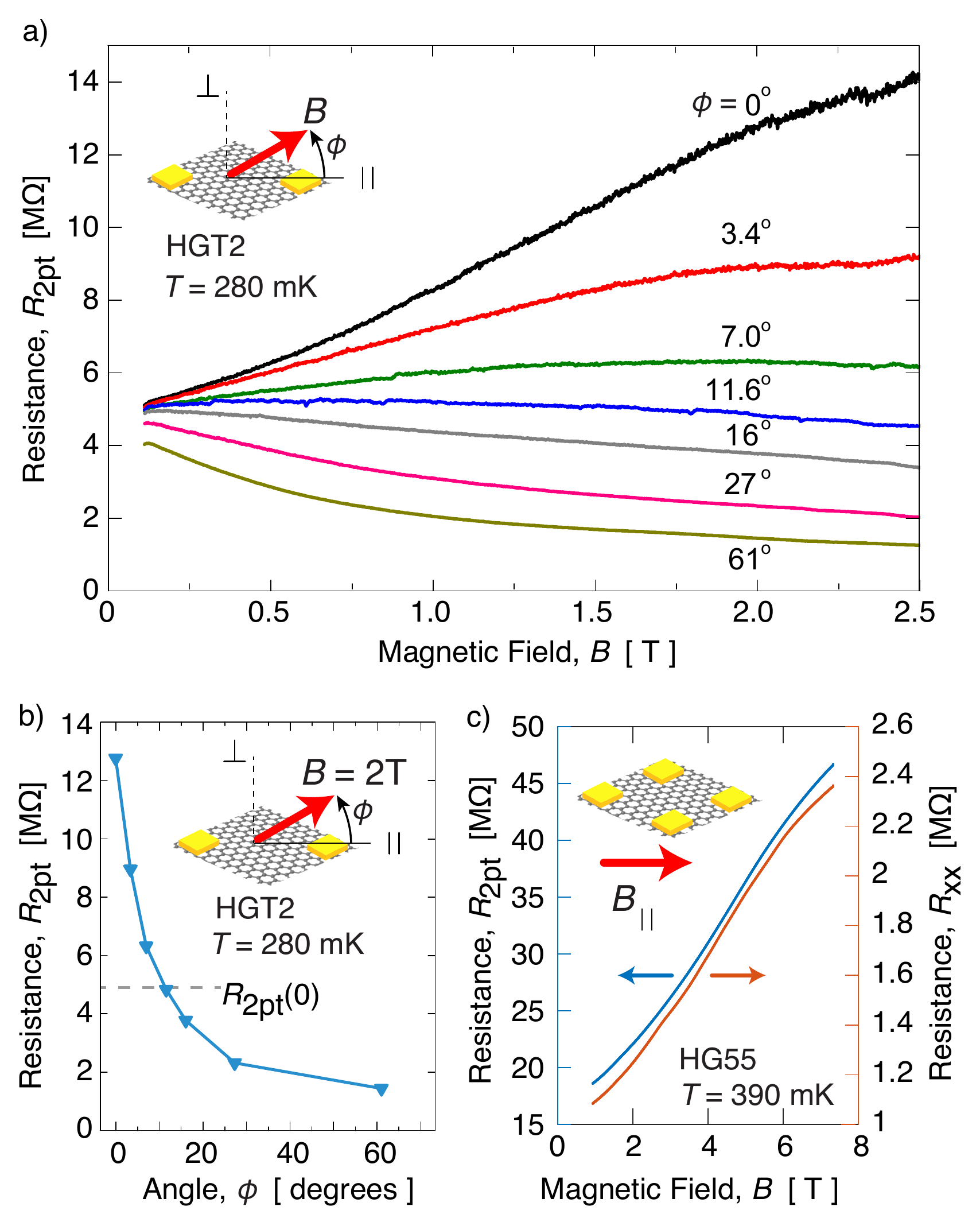} 
		\caption{a) The measured 2-point resistance $R_{\mathrm{2pt}}$ of hydrogenated graphene sample HGT2 versus magnetic field $B$ at a temperature $T=280~\mathrm{mK}$, with different field orientations indicated by the angle $\phi$. Positive MR with an in-plane field gives way to negative MR as the out-of-plane component increases. b) The 2-point resistance $R_{\mathrm{2pt}}$ versus angle $\phi$ at a fixed magnetic field $B = 2.0~\mathrm{T}$, showing the narrow peak in MR versus angle. The zero field resistance $R_{\mathrm{2pt}}(0)$ is indicated. c) A comparison of the measured 2-point resistance $R_{\mathrm{2pt}}$ and 4-point resistance $R_{\mathrm{xx}}$ of hydrogenated graphene sample HG55 versus in-plane magnetic field $B$ at a temperature $T=390~\mathrm{mK}$. Positive MR is observed in both 2-point resistance $R_{\mathrm{2pt}}$ and 4-point resistance $R_{\mathrm{xx}}$. }
	\end{center}
\end{figure}  

To further understand the nature of the magnetoresistance, the angular dependence was investigated with a sample mounted in a $^3$He cryostat in a 5~T split coil superconducting magnet, allowing precise measurement (to $0.1^\circ$) of increments in the angle $\phi$ between applied magnetic field and the graphene plane. The 2-point resistance $R_{\mathrm{2pt}}$ of hydrogenated graphene sample HGT2 versus applied magnetic field $B$ at various angles $\phi$ are illustrated in Fig. 2a) at a temperature of $T=280~\mathrm{mK}$. As the applied field $B$ is rotated out of the in-plane direction $\phi = 0^\circ$, the positive MR is rapidly suppressed and a transition to negative MR occurs at an angle $7^\circ < \phi <12^\circ$ for the magnetic field range investigated, as shown in Fig. 2b). In other words, as the out-of-plane magnetic field component $B_\bot = B \sin (\phi)$ increases, there is a transition from positive to negative MR.

Finally, we confirmed that the positive MR is a bulk effect in the hydrogenated graphene rather than a contact effect alone. Experiments were conducted in a $^3$He cryostat with an 8~T superconducting solenoid, and the sample mounted with the magnetic field in-plane. A comparison of 2-point resistance $R_{\mathrm{2pt}}$ and 4-point resistance $R_{xx}$ versus applied, in-plane magnetic field $B$ is shown in Fig. 2c) for a third sample HG55 with multiple contacts. Care was taken to use a high input impedance ( $Z_{in} \sim 1~\mathrm{T}\Omega$ ) amplifier for 4-point measurement of the highly resistive sample. The 2-point and 4-point resistances both show a large, positive MR, experimentally confirming that the bulk resistivity of hydrogenated graphene exhibits a positive LMR independent of the contact resistance.

We turn our attention to the underlying mechanism for positive MR in hydrogenated graphene with an in-plane field. It is instructive to compare several energy scales and length scales of the problem, with a representative example provided by the observation of positive MR in excess of $100\%$ at $B = 5~\mathrm{T}$ and $T = 320~\mathrm{K}$ in sample HG18, as seen in Fig. 1c). In these experimental conditions, the magnetic length $\ell_B = 11.6~\mathrm{nm}$ at $B = 5~\mathrm{T}$ is far greater than the effective 2DES thickness of $\sqrt{< z^2 >}=0.3~\mathrm{nm}$ in graphene. Thus, magneto-orbital effects are perturbative, and small, in agreement with the absence of a measurable in-plane MR in pristine graphene \cite{chiappini16}. In contrast, the polarization of electron spins is significant. The Zeeman splitting at $B=5~\mathrm{T}$ is $g \mu_B B = 580~\mathrm{\mu eV}$, which is significantly larger than the thermal energy $k_B T = 27.6~\mathrm{\mu eV}$ at $T = 320~\mathrm{mK}$. In other words, $g \mu B / k_B T \gg 1$, and the spin degree of freedom of localized moments is strongly polarized at the $B/T$ ratios for which in-plane LMR is observed. Lastly, the strongly insulating behaviour $\partial R_{\mathrm{2pt}} / \partial T < 0$ of hydrogenated graphene directly implies the onset of electron localization \cite{guillemette13,bennaceur}. Electron transport proceeds by hopping conduction, rather than band conduction.

Our observations are qualitatively consistent with the seminal theoretical work of Kamimura \textit{et al.}\cite{kamimura}, who was the first to show that the combination of spin polarization and electron-electron interaction can lead to strong positive MR in a localized electron system. Briefly, conduction in a localized system can proceed via one of four hopping processes involving unoccupied, singly occupied and doubly occupied localized states, as illustrated in Fig. 3a). In a disordered system with on-site potential $U$, unoccupied, singly occupied and doubly occupied localized states will all be present at energies in the vicinity of the chemical potential $\mu$, and all states are thus available to participate in electron transport by hopping conduction. The exchange interaction is negligible for a pair of singly occupied sites, $(1,1)$, due to the negligible overlap of electron orbitals. This favours a triplet ground state in a magnetic field $B$ as shown in Fig. 3b). In contrast, a doubly occupied site neighbouring an empty site, $(2,0)$, has a non-negligible exchange energy $J$, favouring a singlet ground state as shown in Fig. 3b). Spin polarization with an applied magnetic field thus suppresses electron hopping from a singly occupied site to a singly occupied site $(1,1) \rightarrow (0,2)$, by Pauli blockade, as shown in Fig. 3c). Detailed balance necessarily implies that the hopping process $(2,0) \rightarrow (1,1)$ is also suppressed. The net effect of spin polarization is an attenuation of hopping conduction through a sub-set of paths by the Pauli blockade mechanism, which leads to a suppression of conductance and hence a positive MR.

\begin{figure}%[htbp]
%	\begin{center}
		\includegraphics[width=0.45\textwidth]{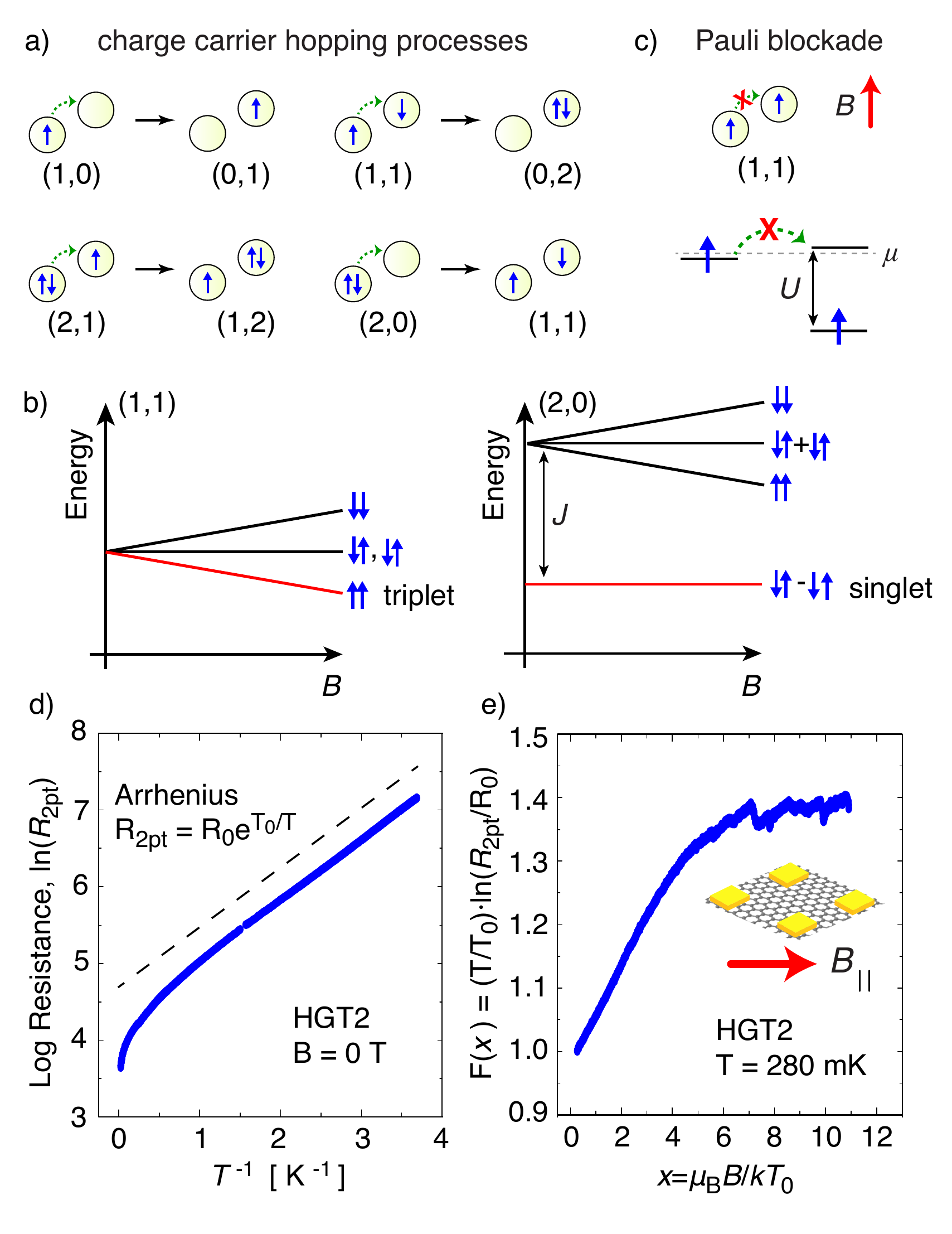} 
		\caption{a) Schematic of four hopping processes between states with intra-state interaction: $(1,0) \rightarrow (0,1)$, a hop from a singly occupied state to an unoccupied state; $(1,1) \rightarrow (0,2)$ a hop from a singly occupied state to a singly occupied state; $(2,1) \rightarrow (1,2)$ a hop from a doubly occupied state to a singly occupied state; and $(2,0) \rightarrow (1,1)$ a hop from a doubly occupied state to an unoccupied state. b) The $(1,1)$ ground state is a triplet, while exchange interaction $J$ favours a singlet ground state for $(2,0)$. c) As a consequence, spin polarization with a magnetic field suppresses the $(1,1) \rightarrow (0,2)$ hopping process by Pauli blockade. $U$ is the on-site Coulomb interaction. d) Measured resistance in natural logarithmic units versus reciprocal temperature of HGT2, showing an Arrhenius thermal activation for $T<1$~K. e) The exponent of magnetoresistance, $F(x)$, versus the dimensionless parameter $x = \mu_B B / k T_0$, determined from the in-plane MR of HGT2 at $T = 280$~mK.  }
		\label{fig:Setup}
%	\end{center}
\end{figure}

A detailed quantitative theory of positive MR was developed by Matveev \textit{et al.} \cite{matveev95} for localized systems that exhibit Mott variable range hopping conduction, and where cyclotron motion could be neglected. The model of Matveev \textit{et al.} assumes a \textit{constant} density of localized states, thus leading to a temperature dependent conductivity $\sigma_{xx} = \sigma_0 \exp \left[ -(T/T_0) ^ {1/{d+1}} F(x) \right] $, where $d = 2,3$ is the dimension of the conductor, $T_0$ is the characteristic Mott temperature, and $F(x)$ is a universal function of a dimensionless magnetic field parameter $x = \mu_B B / k_B T (T_0/T)^{1/{d+1}}$. Experimentally measured hydrogenated graphene resistivity does \textit{not} follow a simple Mott variable range hopping law \cite{guillemette13, bennaceur, hemsworth}  (see Supplemental Material \cite{SM}). The measured resistance of HGT2 in logarithmic scale versus reciprocal temperature $1/T$ at zero magnetic field is shown in Fig. 3d), where agreement with an Arrhenius $R = R_0 \exp(T_0/T)$ thermal activation law is observed for $T<1~\mathrm{K}$. The activation temperature $T_0 = 0.785~\mathrm{K}$ is found by linear fit for $T<1~\mathrm{K}$. Arrhenius activated charge transport is an indication of a well defined energy barrier for hopping, as opposed to the broad distribution of energy barriers appropriate to Mott variable range hopping.

For comparative purposes, we analyze the resistance versus magnetic field using the analytical form derived in Matveev \textit{et al.} \cite{matveev95} appropriate to an Arrhenius activated law, wherein  $R = R_0 \exp[ T_0/T\cdot F(x) ]$ and $x = \mu_B B / k_B T_0$. The factor $F(x)$ for magnetoresistance can be extracted from measured magnetoresistance as $F(x) = (T/T_0)\cdot \ln (R/R_0)$, and is shown versus $x$ for sample HGT2 in Fig. 3e). Numerical computations of in-plane magnetoresistance and $F(x)$ for hydrogenated graphene are presently unavailable. Our experimentally determined $F(x)$ nonetheless adheres to two predictions in the analysis of Matveev \textit{et al.} \cite{matveev95}: $F(x)$ has the linear form $F(x) \approx 1 + k x$ for $x << 1$,  and $F(x) \rightarrow \mathrm{constant}$ for $x>x_{th}$, where $k$ is a constant and $x_{th}$ is a threshold for saturated magnetoresistance.

We now turn our attention to the tilted field regime. The strong suppression of positive MR with an out-of-plane magnetic field component can be understood within this theoretical model. Spin polarization grows with the total field $B$, and thus the suppression of a sub-set of hopping conduction paths grows with the total applied field $B$. The positive MR effect evidently saturates, as expected from the saturation of spin polarization \cite{kamimura, matveev95}, and as directly observed in our experiments (see Fig. 1c)). In contrast, the out-of-plane component $B_\bot = B \sin(\phi)$ induces in-plane cyclotron motion, suppressing back-scattering to produce an overall enhancement of conduction. In the extreme limit, a transition from an insulating state to a quantum Hall state is observed at high out-of-plane field \cite{guillemette13}. Thus, experiments with out-of-plane magnetic field conclusively demonstrate that the enhancement of conduction (negative MR) by in-plane cyclotron motion overwhelms the the suppression of conduction (positive MR) by Pauli blockade. As a consequence, as the applied magnetic field $B$ increases, the positive MR peak versus the angle $\phi$ (between $B$ and the graphene plane) such as that shown in Fig.  2b) is expected to sharpen around $\phi = 0^\circ$. The narrow positive MR peak versus $\phi$ is thus a manifestation of the extreme anisotropy inherent to an atomically thin 2DES.

Our discovery of positive, in-plane LMR in hydrogenated graphene is an experimental observation of the role of spin polarization on hopping conduction in a localized 2DES. The LMR appears at cryogenic temperatures alone where significant spin polarization can be developed by the applied in-plane field. Graphene, and its hydrogenated derivative, is a comparatively benign host for electron spins on account of the weak spin-orbit coupling present. Nonetheless, spin can play a dominant role in MR even in such a benign host due to the Pauli blockade mechanism. The in-plane MR of atomic monolayers and van der Waals heterostructures remains an under-explored area of research, while the development of atomically thin electronics exhibiting LMR is expected to be of interest for ultra-compact magnetic sensing and information storage applications.

\section{Acknowledgements}

The authors thank the Natural Sciences and Engineering Research Council of Canada, Canada Research Chairs program, the Canadian Institute for Advanced Research, the Fonds du Recherche Qu\'eb\'ecois - Natures et Technologies and Hydro-Qu\'ebec for financial support of this work. A portion of this work was performed at the National High Magnetic Field Laboratory which is supported by NSF Cooperative Agreement No. DMR-0084173, the State of Florida, and the DOE.

\clearpage

\section{Supplemental Material}

\subsection{Temperature dependence of resistance}
The resistance versus temperature of the hydrogenated graphene samples in this study do not follow the Mott variable range hopping (VRH) law for a 2-dimensional conductor, nor does the resistance follow the Efros-Shklovskii variable range hopping (ES VRH) law. The former corresponds to a functional form $R = R_0 \exp[ (T_0/T)^{1/3} ]$ and the latter corresponds to a functional form $R = R_0 \exp[ (T_0/T)^{1/2} ]$. We consider sample HGT2 here. A plot of $\ln(R)$ versus $T^{-1/3}$ is shown in Fig. 4 a), and a plot of $\ln(R)$ versus $T^{-1/2}$ is shown in Fig. 4 b). The measured resistance does not follow a linear relation in either plot, indicating that neither a Mott variable range hopping law nor an Efros-Shklovskii variable range hopping law describes the measured resistance.

\begin{figure}[b]
\includegraphics[width=0.45\textwidth]{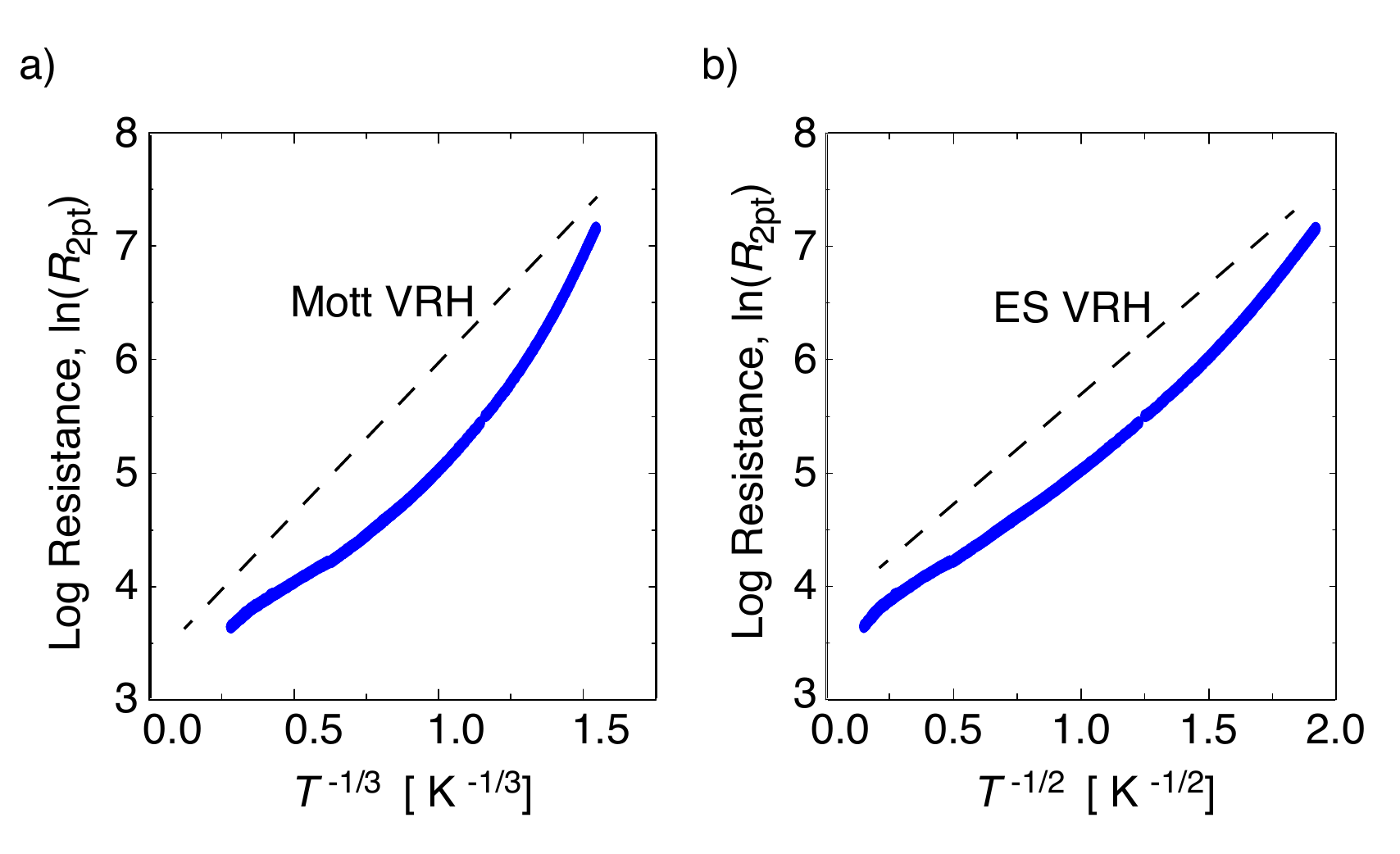}
\caption{(a) A plot of $\ln(R)$ versus $T^{-1/3}$ and (b) a plot of $\ln(R)$ versus $T^{-1/2}$ in zero magnetic field for sample HGT2. }
\end{figure}

\end{document}